\documentclass[conference]{IEEEtran}

\IEEEoverridecommandlockouts

\usepackage{cite}

\usepackage[cmex10]{amsmath}
\usepackage{graphicx}

\DeclareMathOperator{\Prob}{P\!}
\DeclareMathOperator{\Qfunc}{Q\!}
\def\GG{_\textnormal{G}}
\def\BB{_\textnormal{B}}
\def\diff{_\textnormal{diff}}
\def\subs{_{n=\hat{n},T=\hat{T}}}

\begin{document}
\title{Optimizing Pilot Length for a Go/No-Go Decision in Two-State Block Fading Channels with Feedback}

\author{\IEEEauthorblockN{Chung-Yu Lou, Babak Daneshrad, and Richard D. Wesel}
\IEEEauthorblockA{Department of Electrical Engineering, University of California, Los Angeles, California 90095--1594\\
Email: chungyulou@ucla.edu, \{babak, wesel\}@ee.ucla.edu}
\thanks{This material is based upon work supported by the National Science Foundation under Grant 1162501. Any opinions, findings, and conclusions or recommendations expressed in this material are those of the author(s) and do not necessarily reflect the views of the National Science Foundation.}
}


\maketitle

\begin{abstract}
We propose an approach where each user independently seeks to minimize the amount of time that they occupy the channel. Essentially, we seek to minimize the number of transmitted symbols required to communicate a packet assuming variable-length coding with feedback. Users send a pilot sequence to estimate the channel quality and decide whether to proceed with a transmission or wait for the next opportunity. Thus a user may choose to leave the channel even though it has already gained access, in order to increase the network throughput and also save its own energy resources. This paper optimizes the number of pilots and the channel identification threshold to minimize the total number of transmitted symbols (including pilots) required to communicate the packet. We prove a sufficient condition for the optimal pilot length and the channel identification threshold. This optimal parameter pair is solved numerically and the reduction in channel occupancy is shown for various channel settings.

\end{abstract}


\IEEEpeerreviewmaketitle

\section{Introduction}
This paper considers a multi-user wireless network in which each user communicates packets using variable-length coding with feedback. This setting is general in that it includes a range of techniques from simply repeating packets (with or without Chase combining \cite{Chase_Combining}) based on ACKs and NACKs to schemes with decoding attempts after every received symbol with potentially frequent feedback to the transmitter. Our interest lies in a distributed network where global channel state information (CSI) is not available to the users and each user must learn their own channel state. Furthermore, we consider applications such as a sensor network where resource management may take precedence over fairness concerns.  

Techniques using feedback such as the incremental redundancy (IR) scheme proposed in \cite{Incremental_Redundancy} have long been used to enhance link robustness. Recently, IR combined with hybrid automatic repeat request (HARQ) has gained significant attention \cite{IR_HARQ_1,IR_HARQ_2,IR_HARQ_Old_CSI,Adam_Limited_IR,RCSP_Feedback}. In addition to improving the robustness of a link, feedback can also be used to inform the transmitter that the channel is bad enough that the communication attempt should be discontinued until a fade abates.  Rather than struggling in a deep fade or extremely noisy channel, postponing transmission until the channel improves releases the channel for other users in a multi-user setting.

In a distributed multi-user wireless network, the multiple access protocol is in general based on fairness, such as slotted ALOHA or carrier sense multiple access with collision avoidance (CSMA/CA). In this case, the network throughput is reduced by a pair of transmitter (TX) and receiver (RX) with low signal-to-noise ratio (SNR) accessing the channel. Overall network utility can be improved by sacrificing some fairness to include the realities of the channel quality by prioritizing the users in part based on their SNRs. 

However, the SNR is usually not known before sending the pilots. Therefore, we consider a mechanism whereby a user can choose to leave the channel once it believes its SNR is below a certain threshold. When a TX starts to access the channel, it first sends pilots to estimate the SNR, and the RX will respond whether the data transmission should be attempted. If the user decides to leave the channel, it will wait for at least a channel coherence time. By doing so, the user can reduce its energy consumption while also increasing the total network throughput.

An accurate channel estimation relies on sufficient training pilots. However, sending too many pilots unacceptably lowers the rate of that transmission. This trade-off is discussed in \cite{Training_Data_Sum_Limit_1,Training_Data_Sum_Limit_2}, where the total duration of training and data transmission is bounded by the channel coherence time. In \cite{Training_Feedback_Data_Sum_Limit_1,Training_Feedback_Data_Sum_Limit_2}, non-ideal CSI feedback is considered and the duration of the RX sending the CSI back to the TX is also taken into account. In this case, the total duration of training, CSI feedback, and data transmission is limited by the channel coherence time.

In this work, we consider a two-state block fading channel where the channel coefficient remains constant during the training and data transmission. Because of the IR scheme, even though the message length is constant, the transmitted packet has a variable length because the duration required to transmit successfully depends on the SNR. Therefore, we will minimize the number of transmitted symbols (the channel usage) required by a packet transmission, which includes both the initial and any subsequent pilot sequences and the variable length of the data transmission itself. Thus by optimizing the pilot length and the threshold to leave the channel, the user is capable of smart channel selection with a relatively low cost of training. We prove that the optimal parameter pair can be solved numerically and provide a procedure to compute it. The channel usage reduction of avoiding deep fades is shown for various channel settings.

The paper is organized as follows:  Section~\ref{sec:System_Model} describes the system model.  Section~\ref{sec:Main_Proof} proves the optimality of our procedure for computing the optimal parameters for the pilot length and the threshold to leave the channel. Section~\ref{sec:Numerical_Results} shows the reduction in channel occupancy achieved by avoiding bad channels through training. Section~\ref{sec:Conclusion} concludes this paper.

\section{System Model}
\label{sec:System_Model}
When the TX starts to access the channel, it first sends $n$ training symbols to estimate the SNR of the block fading channel for this block. In a block fading channel, the training symbols as well as the data symbols in the same block experience the same fading coefficient.  This coefficient is independent of all the previous blocks. In this paper, we consider a two-state Gaussian channel, i.e. the fading coefficient has two possible realizations, and the received symbols are impacted by additive white Gaussian noise (AWGN). 

We assume that the RX knows the noise power by monitoring the background noise while no signal is present, and the received signal is properly scaled such that the noise has unit power.  Let the $i^\textnormal{th}$ received training symbol be
\begin{equation}
Y_i=X+N_i
\label{eqn:Received_Pilot},
\end{equation}
where $X$ is a positive random variable which depends on the channel fading coefficient and $N_i\sim\mathcal{N}\left(0,1\right)$ is the noise. Let the two possible realizations of $X$ be $x\GG$ and $x\BB$ with probabilities $p\GG$ and $p\BB$, where $p\GG+p\BB=1$, representing the good and bad states of the channel, respectively. In other words, the SNRs of the good and bad states are $x\GG^2$ and $x\BB^2$, where $x\GG>x\BB$. The statistics of the channel $\left(p\GG,p\BB,x\GG,x\BB\right)$ are known, and the realization of $X$ is to be estimated. Note that, if either $p\GG=0$, $p\BB=0$, or $x\GG=x\BB$, then this two-state channel is essentially a static channel and needs no training at all. Therefore, we only consider the non-trivial cases.

Having $n$ measures of the random variable $X$ affected by the Gaussian noise, the RX then uses the sufficient statistic
\begin{equation}
\bar{Y}=\frac{1}{n}\sum\limits_{i=1}^n{Y_i}
\label{eqn:Averaged_Received_Pilot}
\end{equation}
to determine whether the data transmission should begin by comparing it with a threshold $T$. If $\bar{Y}\geq T$, the RX believes that the channel is in the good state and allows the TX to initiate the data transmission. Otherwise, the RX asks the TX to leave the channel and wait for the next block.  On the next block that the TX gains access to the channel, the training process starts over again because the channel realization in the new block is independent of the blocks observed in previous attempts. 

In this study, we assume for simplicity that the feedback is noiseless and instantaneous, but note that this feedback is of a single bit of information and is required only once per transmission attempt so that this assumption neglects a small fixed overhead.  Since $\bar{Y}\sim\mathcal{N}\left(X,\frac{1}{n}\right)$, the conditional probabilities of initiating the data transmission while the channel is in the good and bad states are given by
\begin{IEEEeqnarray*}{rCl}
\Prob\left(\bar{Y}\geq T\middle|X=x\GG\right)&=&\Qfunc\left(\sqrt{n}\left(T-x\GG\right)\right)\\
\Prob\left(\bar{Y}\geq T\middle|X=x\BB\right)&=&\Qfunc\left(\sqrt{n}\left(T-x\BB\right)\right),
\end{IEEEeqnarray*}
where $\Qfunc\left(\cdot\right)$ is the Gaussian Q-function
\begin{equation*}
\Qfunc\left(x\right)=\int\limits_x^\infty{\frac{1}{\sqrt{2\pi}}\exp\!\left(\frac{t^2}{-2}\right)\mathrm{d}t}.
\end{equation*}

During the data transmission, the TX sends a packet containing $k$ bits. With the IR scheme, the TX keeps sending data symbols until the RX decodes the packet successfully and sends an acknowledgement. Let the expected channel usage of the data transmission given the channel being in the good and bad states be $\tau\GG$ and $\tau\BB$, respectively, where $0<\tau\GG<\tau\BB$. We denote $\bar{\tau}$ as the total expected channel usage to deliver a packet. This includes one or more length-$n$ pilot sequences. Hence, $\bar{\tau}$ is calculated as
\begin{IEEEeqnarray}{rCl}
\bar{\tau}&=&n+\Prob\left(\bar{Y}\geq T\middle|X\!=x\GG\right)p\GG\tau\GG\nonumber\\
&&+\Prob\left(\bar{Y}\geq T\middle|X\!=x\BB\right)p\BB\tau\BB+\Prob\left(\bar{Y}<T\right)\bar{\tau}
\label{eqn:Expected_Total_Channel_Usage_Definition}.
\end{IEEEeqnarray}
Subtracting the last term of \eqref{eqn:Expected_Total_Channel_Usage_Definition} from both sides produces
\begin{equation}
\bar{\tau}=\frac{n+p\BB\Qfunc\left(\sqrt{n}\left(T-x\BB\right)\right)\tau\diff}{p\GG\Qfunc\left(\sqrt{n}\left(T-x\GG\right)\right)+p\BB\Qfunc\left(\sqrt{n}\left(T-x\BB\right)\right)}+\tau\GG
\label{eqn:Expected_Total_Channel_Usage},
\end{equation}
where $\tau\diff=\tau\BB-\tau\GG>0$.

Our goal is to minimize $\bar{\tau}$ by picking the optimal parameter pair $\left(n^\star,T^\star\right)$. In Section~\ref{sec:Numerical_Results}, we will compare $\left.\bar{\tau}\right|_{n=n^\star,T=T^\star}$ with $\bar{\tau}_\textnormal{ref}$. The referenced channel usage $\bar{\tau}_\textnormal{ref}$ is defined as the expected channel usage to deliver a packet without training and given by
\begin{equation}
\bar{\tau}_\textnormal{ref}=\left.\bar{\tau}\right|_{n=0}=p\GG\tau\GG+p\BB\tau\BB
\label{eqn:Expected_Total_Channel_Usage_Ref}.
\end{equation}
In \eqref{eqn:Expected_Total_Channel_Usage}, it is clear that $\tau\GG$ is a lower bound for $\bar{\tau}$.  Achieving this lower bound requires identifying the low-SNR channel perfectly $\left(\Prob\left(\bar{Y}\geq T\middle|X=x\BB\right)=0\right)$ with absolutely no cost $\left(n=0\right)$. Given the lower bound for $\bar{\tau}=\tau\GG$, we further define a corresponding upper bound on the expected reduction in channel occupancy as
\begin{equation}
\Delta\bar{\tau}=\bar{\tau}_\textnormal{ref}-\tau\GG=p\BB\tau\diff
\label{eqn:Potential_Reduction}.
\end{equation}

\section{Optimal Pilot Length and Threshold}
\label{sec:Main_Proof}
Although our target $n$ is a non-negative integer, \eqref{eqn:Expected_Total_Channel_Usage} can be evaluated for any non-negative real $n$. Thus, to simplify the derivation, our procedure identifies the optimal integer $n^\star$ by examining the integers neighboring the real number $\hat{n}$ that minimizes $\bar{\tau}$.  The selected integer is the optimal integer $n^\star$ because of the continuity of \eqref{eqn:Expected_Total_Channel_Usage}.

The discussion proceeds as follows:   A sufficient condition for a local minimum facilitates a proof that there is only one local minimum, which is the global minimum.  Then, solving numerically for the optimal nonnegative real $\hat{n}$  leads to the optimal integer $n^\star$ and optimal pair $\left(n^\star,T^\star\right)$.

\subsection{Sufficient Condition for a Local Minimum}
Considering the following extreme points establishes that there are no minima on the boundaries:
\begin{itemize}
\item As $n\to\infty$, $\bar{\tau}\to\infty$ and cannot be a local minimum.
\item As $n>0$ and $T\to\infty$, we have $\bar{\tau}\to\infty$ and it cannot be a local minimum.
\item When $n>0$ and $T\to-\infty$, we can shown that $\bar{\tau}=n+\bar{\tau}_\textnormal{ref}$, which is not a local minimum since $n$ can be reduced.
\item When $n=0$, we know $\bar{\tau}=\bar{\tau}_\textnormal{ref}$. If $T$ is finite, then
\begin{equation*}
\left.\frac{\partial}{\partial n}\bar{\tau}\right|_{n=0}=\left.2p\GG p\BB\tau\diff\left(1-\frac{x\GG-x\BB}{2\sqrt{2\pi n}}\right)\right|_{n=0}<0.
\end{equation*}
So local minima do not appear on $n=0$.
\end{itemize}

Thus the only possible local minima occur in the interior of the domain.  Let the gradient and Hessian of $\bar{\tau}$ with respect to $\left(T,n\right)$ be
\begin{equation}
\nabla\bar{\tau}=\begin{bmatrix}
\frac{\partial}{\partial T}\bar{\tau}\\[0.3em]
\frac{\partial}{\partial n}\bar{\tau}
\end{bmatrix}
\label{eqn:Gradient_Definition}
\end{equation}
and
\begin{equation}
\nabla^2\bar{\tau} = \begin{bmatrix}
\frac{\partial^2}{\partial T^2}\bar{\tau} & \frac{\partial^2}{\partial T\partial n}\bar{\tau}\\[0.3em]
\frac{\partial}{\partial n\partial T}\bar{\tau} & \frac{\partial^2}{\partial n^2}\bar{\tau}
\end{bmatrix}
\label{eqn:Hessian_Definition}.
\end{equation} 
Because the gradient and Hessian are continuous, a local minimum satisfies that its gradient is a zero vector, and its Hessian is positive definite.  

Let $\hat{n}$ and $\hat{T}$ be a solution of $\nabla\bar{\tau}=0$. We will derive a relationship between $\hat{n}$ and $\hat{T}$ that shows for any such $\hat{n}$ and $\hat{T}$, $\left.\nabla^2\bar{\tau}\right|\subs>0$. 

By evaluating $\left.\frac{\partial}{\partial T}\bar{\tau}\right|\subs=0$, we obtain
\begin{IEEEeqnarray}{l}
\left.\left(p\GG Q\GG+p\BB Q\BB\right)\!\left(p\BB\frac{\partial}{\partial T}Q\BB\tau\diff\right)\right|\subs\nonumber\\
\quad=\!\left.\left(n+p\BB Q\BB\tau\diff\right)\!\left(p\GG\frac{\partial}{\partial T}Q\GG+p\BB\frac{\partial}{\partial T}Q\BB\right)\right|\subs\!
\label{eqn:Partial_T},
\IEEEeqnarraynumspace
\end{IEEEeqnarray}
where we define $Q\GG=\Qfunc\left(\sqrt{n}\left(T-x\GG\right)\right)$, and $Q\BB=\Qfunc\left(\sqrt{n}\left(T-x\BB\right)\right)$. Similarly, from $\left.\frac{\partial}{\partial n}\bar{\tau}\right|\subs=0$, we get
\begin{IEEEeqnarray}{l}
\left.\left(p\GG Q\GG+p\BB Q\BB\right)\!\left(1+p\BB\frac{\partial}{\partial n}Q\BB\tau\diff\right)\right|\subs\nonumber\\
\quad=\left.\left(n+p\BB Q\BB\tau\diff\right)\!\left(p\GG\frac{\partial}{\partial n}Q\GG+p\BB\frac{\partial}{\partial n}Q\BB\right)\right|\subs
\label{eqn:Partial_n}.
\IEEEeqnarraynumspace
\end{IEEEeqnarray}

In order to show that $\left.\nabla^2\bar{\tau}\right|\subs$ is positive definite, we need to prove the two inequalities
\begin{equation}
\left.\frac{\partial^2}{\partial T^2}\hat{\tau}\right|\subs>0
\label{eqn:Inequality_1}
\end{equation}
and
\begin{equation}
\left.\det\nabla^2\bar{\tau}\right|\subs>0
\label{eqn:Inequality_2}.
\end{equation}
Using \eqref{eqn:Partial_T} and the fact that $\left.n+p\BB Q\BB\tau\diff\right|\subs>0$ and $\left.p\BB\frac{\partial}{\partial T}Q\BB\tau\diff\right|\subs<0$, the first inequality \eqref{eqn:Inequality_1} is equivalent to
\begin{IEEEeqnarray*}{l}
\left.\left(p\GG\frac{\partial}{\partial T}Q\GG+p\BB\frac{\partial}{\partial T}Q\BB\right)\left(p\BB\frac{\partial^2}{\partial T^2}Q\BB\tau\diff\right)\right|\subs\\
\quad<\left.\left(p\BB\frac{\partial}{\partial T}Q\BB\tau\diff\right)\left(p\GG\frac{\partial^2}{\partial T^2}Q\GG+p\BB\frac{\partial^2}{\partial T^2}Q\BB\right)\right|\subs.
\end{IEEEeqnarray*}
After evaluating the partial derivatives of the above expression, \eqref{eqn:Inequality_1} is also equivalent to 
\begin{equation}
\hat{n}^{\frac{3}{2}}p\GG p\BB \exp\!\bigg(\frac{\hat{G}^2+\hat{B}^2}{-2}\bigg)(\hat{B}-\hat{G})>0
\label{eqn:Partial_T2},
\end{equation}
where
\begin{IEEEeqnarray}{rCl}
\hat{G}&=&\sqrt{\hat{n}}\,(\hat{T}-x\GG)
\label{eqn:Definition_G_Hat}\\
\hat{B}&=&\sqrt{\hat{n}}\,(\hat{T}-x\BB)
\label{eqn:Definition_B_Hat}.
\end{IEEEeqnarray}
Furthermore, \eqref{eqn:Partial_T2} is always true because every term on the left-hand side (LHS) is positive, including $\hat{B}-\hat{G}=\sqrt{n}\left(x\GG-x\BB\right)>0$.

Regarding the second inequality \eqref{eqn:Inequality_2}, we use \eqref{eqn:Partial_T} and \eqref{eqn:Partial_n} multiple times and the fact that both $\left.n+p\BB Q\BB\tau\diff\right|\subs$ and $\left.p\BB\frac{\partial}{\partial T}Q\BB\tau\diff\right|\subs$ are non-zero to change the Q-functions and their first order partial derivatives with respect to $n$ into the first order partial derivatives with respect to $T$. The inequality \eqref{eqn:Inequality_2} simplifies to
\begin{IEEEeqnarray*}{l}
\left.\left(\frac{\partial}{\partial T}Q\GG\frac{\partial^2}{\partial T^2}Q\BB-\frac{\partial}{\partial T}Q\BB\frac{\partial^2}{\partial T^2}Q\GG\right)\right|\subs\\
\left.\cdot\left(\frac{\partial}{\partial T}Q\GG\frac{\partial^2}{\partial n^2}Q\BB-\frac{\partial}{\partial T}Q\BB\frac{\partial^2}{\partial n^2}Q\GG\right)\right|\subs\\
\qquad>\!\left.\left(\frac{\partial}{\partial T}Q\GG\frac{\partial^2}{\partial n\partial T}Q\BB-\frac{\partial}{\partial T}Q\BB\frac{\partial^2}{\partial n\partial T}Q\GG\right)^2\right|\subs.
\end{IEEEeqnarray*}
After evaluating the partial derivatives, the above inequality is equivalent to 
\begin{equation*}
(\hat{B}-\hat{G})\big[\hat{B}\,\big(1+\hat{B}^2\big)\!-\hat{G}\,\big(1+\hat{G}^2\big)\big]>\big(\hat{B}^2-\hat{G}^2\big)^2,
\end{equation*}
which is equivalent to
\begin{equation}
1-\hat{G}\hat{B}>0
\label{eqn:det}.
\end{equation}

To prove \eqref{eqn:det}, we must rely on a relationship between $\hat{n}$ and $\hat{T}$. Specifically, from \eqref{eqn:Partial_T}, we get
\small\begin{IEEEeqnarray}{l}
\hat{n}\left(p\GG\,e^{\frac{\hat{G}^2}{-2}}+p\BB\,e^{\frac{\hat{B}^2}{-2}}\right)=p\GG p\BB\tau\diff\!\left[\Qfunc\,(\hat{G})\,e^{\frac{\hat{B}^2}{-2}}-\Qfunc\,(\hat{B})\,e^{\frac{\hat{G}^2}{-2}}\right]\!
\label{eqn:Partial_T_Simplify}.
\IEEEeqnarraynumspace
\end{IEEEeqnarray}\normalsize
Multiplying \eqref{eqn:Partial_n} with $\!\left.\left(p\BB\frac{\partial}{\partial T}Q\BB\tau\diff\right)\!\middle/\!\left(n\!+\!p\BB Q\BB\tau\diff\right)\right|\!\subs$ and then using \eqref{eqn:Partial_T}, we know
\begin{equation}
\hat{n}\left(p\GG\,e^{\frac{\hat{G}^2}{-2}}+p\BB\,e^{\frac{\hat{B}^2}{-2}}\right)=p\GG p\BB\tau\diff\,e^{\frac{\hat{G}^2+\hat{B}^2}{-2}}\frac{\hat{G}-\hat{B}}{-2\sqrt{2\pi}}
\label{eqn:Partial_n_Simplify}.
\end{equation}
Since \eqref{eqn:Partial_T_Simplify} and \eqref{eqn:Partial_n_Simplify} have the same LHS, we obtain the equality
\begin{equation}
\hat{G}+2\sqrt{2\pi}\Qfunc\,(\hat{G})\,e^{\frac{\hat{G}^2}{2}}=\hat{B}+2\sqrt{2\pi}\Qfunc\,(\hat{B})\,e^{\frac{\hat{B}^2}{2}}
\label{eqn:Partial_nT_Simplify},
\end{equation}
where $\hat{G}$ only appears on the LHS and $\hat{B}$ only appears on the right-hand side (RHS). By introducing the function
\begin{equation}
f\!\left(x\right)=x+2\sqrt{2\pi}\Qfunc\left(x\right)\exp\!\left(\frac{x^2}{2}\right)
\label{eqn:Mysterious_Func},
\end{equation}
we get $f(\hat{G})=f(\hat{B})$. The second derivative of this function is given by
\begin{equation}
f''\!\left(x\right)=2\sqrt{2\pi}\Qfunc\left(x\right)\exp\left(\frac{x^2}{2}\right)\left(x^2+1\right)-2x
\label{eqn:Mysterious_Func_Second_Derivative}.
\end{equation}
From \eqref{eqn:Mysterious_Func_Second_Derivative}, $f''\!\left(x\right)>0$ when $x\leq0$. When $x>0$, we can also show $f''\!\left(x\right)>0$ through the Gaussian Q-function lower bound \cite{Q_Function_x_Bound}
\begin{equation*}
\Qfunc\left(x\right)>\frac{x}{x^2+1}\frac{1}{\sqrt{2\pi}}\exp\!\left(\frac{x^2}{-2}\right)\qquad x>0.
\end{equation*}
Hence $f\!\left(x\right)$ is strictly convex.
\begin{figure}[!t]
\centering
\includegraphics[width=0.99\columnwidth]{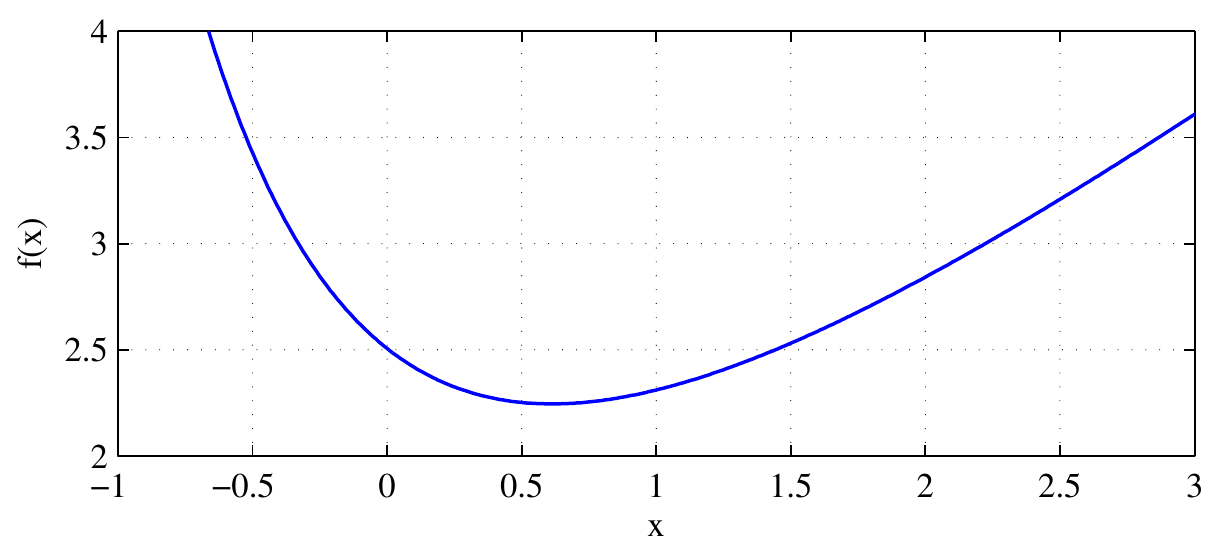}
\caption{The function $f\!\left(x\right)$ given in \eqref{eqn:Mysterious_Func}.}
\label{fig:Mysterious_Function}
\end{figure}
Having $f\!\left(x\right)$ plotted in Fig.~\ref{fig:Mysterious_Function}, we can properly draw a horizontal line and create two intersections such that the $x$-coordinates of the left and right intersections correspond to $\hat{G}$ and $\hat{B}$ respectively, because $f(\hat{G})=f(\hat{B})$ and $\hat{G}<\hat{B}$. Since the minimum of $f\!\left(x\right)$ happens at $x=0.6120$, we have $\hat{G}<0.6120<\hat{B}$. To prove \eqref{eqn:det}, we shall discuss the two cases of $\hat{G}$:
\begin{itemize}
\item If $\hat{G}\leq0$, it is straightforward that $1-\hat{G}\hat{B}>0$ because $\hat{B}>0$.
\item If $0<\hat{G}<0.6120$, then $2.5066<f(\hat{G})<2.2460$, which yields $1.4536>\hat{B}>0.6120$. Combining the ranges of $\hat{G}$ and $\hat{B}$, we have $1-\hat{G}\hat{B}>0.1104>0$.
\end{itemize}
Thus, \eqref{eqn:det} is valid, and it leads to $\left.\det\nabla^2\bar{\tau}\right|\subs>0$.

Since both \eqref{eqn:Inequality_1} and \eqref{eqn:Inequality_2} are true, $\bar{\tau}$ has a local minimum at any point $(\hat{n},\hat{T})$ for which $\left.\nabla\bar{\tau}\right|\subs=0$. We now show that there is exactly one such point.

\subsection{Existence and Uniqueness of the Local Minimum}
From the definition of $\hat{G}$ and $\hat{B}$ in \eqref{eqn:Definition_G_Hat} and \eqref{eqn:Definition_B_Hat}, the corresponding $(\hat{n},\hat{T})$ can be obtained from $(\hat{G},\hat{B})$ through
\begin{equation}
\hat{n}=\bigg(\frac{\hat{G}-\hat{B}}{x\GG-x\BB}\bigg)^2
\label{eqn:Solve_n_hat}
\end{equation}
and
\begin{equation}
\hat{T}=\frac{\hat{G}x\BB-\hat{B}x\GG}{\hat{G}-\hat{B}}
\label{eqn:Solve_T_hat}.
\end{equation}
Thus, it is sufficient to prove the existence and uniqueness of $(\hat{G},\hat{B})$. Using \eqref{eqn:Solve_n_hat} to substitute for  $\hat{n}$ in \eqref{eqn:Partial_n_Simplify} yields
\begin{equation}
(\hat{B}-\hat{G})\left(p\GG\,e^{\frac{\hat{B}^2}{2}}+p\BB\,e^{\frac{\hat{G}^2}{2}}\right)=p\GG p\BB\tau\diff\frac{\left(x\GG-x\BB\right)^2}{2\sqrt{2\pi}}
\label{eqn:Uniqueness_Equation}.
\end{equation}
Note that the LHS of \eqref{eqn:Uniqueness_Equation} is a function of $(\hat{G},\hat{B})$ while the RHS is a positive constant. We can further simplify the LHS through the function $f\!\left(x\right)$ defined in \eqref{eqn:Mysterious_Func}.

Let $y=f\!\left(x\right)$, and the two inverse functions $x=f\GG^{-1}\!\left(y\right)$ and $x=f\BB^{-1}\!\left(y\right)$ for the ranges $x<0.6120$ and $x>0.6120$, respectively. Then $\hat{G}$ and $\hat{B}$ can be expressed in terms of $y$ and the LHS of \eqref{eqn:Uniqueness_Equation} is $h(y)$ which written as
\begin{IEEEeqnarray}{l}
h\!\left(y\right)=\left[f\BB^{-1}\!\left(y\right)-f\GG^{-1}\!\left(y\right)\right]\!\left[p\GG\,e^{\frac{f\BB^{-1}\!\left(y\right)^2}{2}}+p\BB\,e^{\frac{f\GG^{-1}\left(y\right)^2}{2}}\right]\!
\label{eqn:Single_Variable_Function},
\IEEEeqnarraynumspace
\end{IEEEeqnarray}
which is a function of a single variable $y>2.2460$. When $y$ approaches its minimum, $h\!\left(y\right)$ approaches zero. As $y$ goes to infinity, $h\!\left(y\right)$ goes to infinity as well. Therefore, by continuity, $h\!\left(y\right)$ meets the RHS of \eqref{eqn:Uniqueness_Equation} at least once. This proves the existence of $(\hat{G},\hat{B})$.

To prove the uniqueness, we show that $h\!\left(y\right)$ is strictly increasing. Consider its first derivative $h'\!\left(y\right)$, in which we are able to show
\begin{equation*}
\frac{\mathrm{d}}{\mathrm{d}y}\left[f\BB^{-1}\!\left(y\right)-f\GG^{-1}\!\left(y\right)\right]p\GG\exp\!\left(\frac{f\BB^{-1}\left(y\right)^2}{2}\right)>0
\end{equation*}
using $\frac{\mathrm{d}}{\mathrm{d}y}f\BB^{-1}\!\left(y\right)>0>\frac{\mathrm{d}}{\mathrm{d}y}f\GG^{-1}\!\left(y\right)$, $f\BB^{-1}\!\left(y\right)>f\GG^{-1}\!\left(y\right)$, and $f\BB^{-1}\!\left(y\right)>0$. Therefore, $h'\!\left(y\right)$ is lower bounded by
\begin{IEEEeqnarray*}{rCl}
h'\!\left(y\right)&>&\frac{\mathrm{d}}{\mathrm{d}y}\left[f\BB^{-1}\left(y\right)-f\GG^{-1}\left(y\right)\right]p\BB\exp\!\left(\frac{f\GG^{-1}\left(y\right)^2}{2}\right)\\
&=&p\BB\exp\!\left(\frac{f\GG^{-1}\left(y\right)^2}{2}\right)\Bigg\{\frac{\mathrm{d}}{\mathrm{d}y}f\BB^{-1}\!\left(y\right)\\
&&+\left[f\GG^{-1}\!\left(y\right)f\BB^{-1}\!\left(y\right)-1-f\GG^{-1}\!\left(y\right)^2\right]\frac{\mathrm{d}}{\mathrm{d}y}f\GG^{-1}\!\left(y\right)\!\Bigg\}\yesnumber
\label{eqn:Single_Variable_Function_Derivative}.
\IEEEeqnarraynumspace
\end{IEEEeqnarray*}
Since $\frac{\mathrm{d}}{\mathrm{d}y}f\BB^{-1}\!\left(y\right)>0$, $\frac{\mathrm{d}}{\mathrm{d}y}f\GG^{-1}\!\left(y\right)<0$ and $f\GG^{-1}\!\left(y\right)f\BB^{-1}\!\left(y\right)<1$, the RHS of \eqref{eqn:Single_Variable_Function_Derivative} is positive. Hence, $h\!\left(y\right)$ is strictly increasing, and the uniqueness of $(\hat{G},\hat{B})$ is proven.

To summarize, we have shown that there is a unique solution to $\nabla\bar{\tau}=0$, which implies a local minimum. In other words, the solution to $\nabla\bar{\tau}=0$ yields the global minimum. This guarantees that Newton's method will always converge to the global minimum.

\subsection{Numerical Method}
Given all the parameters $\left(p\GG,p\BB,x\GG,x\BB,\tau\GG,\tau\BB\right)$, the point $(\hat{n},\hat{T})$ that yields the global minimum of $\bar{\tau}$ can be computed using Newton's method. Since $\hat{n}$ may not be an integer, we consider its neighbor integers $\hat{n}_1=\lfloor\hat{n}\rfloor$ and $\hat{n}_2=\lceil\hat{n}\rceil$. For $\hat{n}_1$ and $\hat{n}_2$, the corresponding optimal thresholds $\hat{T}_1$ and $\hat{T}_2$ are computed. The final optimal parameter pair $\left(n^\star,T^\star\right)$ is then chosen between $(\hat{n}_1,\hat{T}_1)$ and $(\hat{n}_2,\hat{T}_2)$, whichever achieves the smaller total expected channel usage $\bar{\tau}$. The computation of $\hat{T}_i$, $i\in\left\{1,2\right\}$ is discussed below.

With the integer $n=\hat{n}_i$ fixed, $\bar{\tau}$ is a function only of $T$. Denote this function as $\bar{\tau}_i\!\left(T\right)$, and let its first and second derivatives be $\bar{\tau}_i'\!\left(T\right)$ and $\bar{\tau}_i''\!\left(T\right)$, respectively. When $\hat{n}_i=0$, the value of $\hat{T}_i$ is not important because $\bar{\tau}_i\!\left(T\right)=\bar{\tau}_\textnormal{ref}$ regardless of $T$. For $\hat{n}_i>0$,  for any finite $\hat{T}_i$ satisfying $\bar{\tau}_i'(\hat{T}_i)=0$, we are able to show $\bar{\tau}_i''(\hat{T}_i)>0$, i.e. $\hat{T}_i$ is a local minimum, by following the proof of \eqref{eqn:Inequality_1}.  Furthermore, we know $\bar{\tau}_i'\!\left(T\right)>0$ for $T\gg0$ because $\lim\nolimits_{T\to\infty}\bar{\tau}_i\!\left(T\right)=\infty$. These two facts and the continuity of $\bar{\tau}_i\!\left(T\right)$ and $\bar{\tau}_i'\!\left(T\right)$ imply that, if it exists, $\hat{T}_i$ is unique and thus it yields the global minimum. 

We now discuss the two cases where this finite solution to $\bar{\tau}_i'\!\left(T\right)=0$ does and does not exist. Consider 
\begin{IEEEeqnarray}{l}
\lim\limits_{T\to-\infty}\bar{\tau}_i'\!\left(T\right)=\lim\limits_{T\to-\infty}\frac{\sqrt{\hat{n}_i}}{\sqrt{2\pi}}e^{\frac{\hat{n}_i\left(T-x\BB\right)^2}{-2}}p\BB\,(\hat{n}_i-p\BB\tau\diff)
\label{eqn:Derivative_Approach_Neg_Infty}.
\IEEEeqnarraynumspace
\end{IEEEeqnarray}

{\em Case 1:} If $0<\hat{n}_i<p\BB\tau\diff$, then $\bar{\tau}_i'\!\left(T\right)<0$ for $T\ll0$ since \eqref{eqn:Derivative_Approach_Neg_Infty} approaches zero from below.
In addition, we know $\bar{\tau}_i'\!\left(T\right)>0$ for $T\gg0$. Thus, because of the intermediate value theorem of the continuous function $\bar{\tau}_i'\!\left(T\right)$, the finite solution to $\bar{\tau}_i'\!\left(T\right)=0$ exists and it is $\hat{T}_i$. Moreover, since the local minimum $\hat{T}_i$ is unique, Newton's method is guaranteed to converge to it.

{\em Case 2:} If $\hat{n}_i>p\BB\tau\diff$, then $\bar{\tau}_i'\left(T\right)>0$ for $T\ll0$ because \eqref{eqn:Derivative_Approach_Neg_Infty} approaches zero from above. This implies that there must be an even number of finite roots of $\bar{\tau}_i'\!\left(T\right)=0$. Since $\hat{T}_i$ must be unique if it exists, there is no such root\footnote{Repeated roots are not allowed because $\bar{\tau}_i''(\hat{T}_i)>0$.}. In this case, $\hat{\tau}\!\left(T\right)$ is strictly increasing and the global minimum happens at $T\to-\infty$. Having such threshold means that the user will always initiate the data transmission after the training. This is reasonable because the price of training $\hat{n}_i$ is greater than the upper bound for the expected reduction in channel occupancy $\Delta\bar{\tau}=p\BB\tau\diff$ defined in \eqref{eqn:Potential_Reduction}. Furthermore, if $\hat{n}_i=p\BB\tau\diff$, the same threshold should be used for the same reason.

The procedure to compute the optimal parameter pair $\left(n^\star,T^\star\right)$ is summarized as follows:
\begin{enumerate}
\item Find the optimal parameter pair $(\hat{n},\hat{T})$ that yields the global minimum of $\bar{\tau}$ using Newton's method, where $\hat{n}$ is a positive real number.
\item Let $\hat{n}_1=\lfloor\hat{n}\rfloor$ and $\hat{n}_2=\lceil\hat{n}\rceil$. Calculate $\hat{T}_i$ for $i\in\left\{1,2\right\}$ according to:
\begin{itemize}
\item If $\hat{n}_i=0$, set $\hat{T}_i=0$.
\item If $0<\hat{n}_i<\Delta\bar{\tau}$, find $\hat{T}_i$ that yields the global minimum of $\left.\bar{\tau}\right|_{n=\hat{n}_i}$ using Newton's method.
\item If $\hat{n}_i\geq \Delta\bar{\tau}$, set $\hat{T}_i=-\infty$.  This can only happen when $i=2$, and implies that $\hat{n}_1$ will be optimal. 
\end{itemize}
\item Pick the smaller total expected channel usage between $\left.\bar{\tau}\right|_{n=\hat{n}_1,T=\hat{T}_1}$ and $\left.\bar{\tau}\right|_{n=\hat{n}_2,T=\hat{T}_2}$, and return the corresponding parameter pair.
\end{enumerate}

\section{Numerical Results}
\label{sec:Numerical_Results}
The approach described above to find the optimal parameter pair $\left(n^\star,T^\star\right)$ is general. It can be applied to any transmission scheme that has the essential elements of feedback to let the transmitter know whether to proceed with the communication and also variable-length coding and feedback that would permit transmission with expected lengths of $\tau\GG$ on the good channels and $\tau\BB$ on the bad channels.

This section provides numerical results when the optimal parameter pairs $\left(n^\star,T^\star\right)$ are used with a specific scheme for variable-length coding with feedback. We expect that the qualitative behavior would be similar for many other such schemes.  

We consider a message length of $k=128$ bits, which are encoded using a rate-compatible code into a variable-length packet whose bits are modulated using binary phase-shifted keying (BPSK).  Our example uses a feedback scheme in which the RX tries to decode the variable-length packet after receiving every BPSK symbol and will reply with an acknowledgement once it decodes successfully. 

This IR scheme can be analyzed through rate-compatible sphere-packing (RCSP) \cite{RCSP_Feedback}. The probability of decoding error after receiving $m$ symbols is
\begin{equation}
\Prob_m\!\left(\gamma\right)=1-\mathrm{F}_{\chi^2\left(m\right)}\!\left(\frac{m\left(1+\gamma\right)}{2^{\left.2k\middle/m\right.}}\right)
\label{eqn:RCSP_Error_at_m},
\end{equation}
where $\gamma$ is the SNR and $\mathrm{F}_{\chi^2\left(m\right)}\!\left(\cdot\right)$ is the cumulative distribution function (CDF) of the chi-square distribution with $m$ degrees of freedom. To guarantee a successful data transmission, we do not put any limit on the number of available data symbols, and $\lim\nolimits_{m\to\infty}{\Prob_m\!\left(\gamma\right)}=0$. With the probability upper bound given in \cite{RCSP_Feedback_Bound}, the expected channel usage of the data transmission is upper bounded by
\begin{equation*}
\tau_\textnormal{data}\!\left(\gamma\right)\leq1+\sum\limits_{m=1}^\infty{\Prob_m\!\left(\gamma\right)}.
\end{equation*}

While computing $\tau\GG$ and $\tau\BB$, only finite terms are evaluated. For instance,
\begin{equation*}
\tau\GG\approx 1+\sum\limits_{m=1}^{\hat{m}\GG}{\Prob_m\!\left(x\GG^2\right)},
\end{equation*}
where $\hat{m}\GG$ is the smallest positive integer that satisfies the following two criteria. The first criterion is $\left(1+x\GG^2\right)2^{\left.-2k\middle/\hat{m}\GG\right.}>1+0.5x\GG^2$. Thus the sum of the omitted terms is upper bounded by
\begin{IEEEeqnarray}{rCl}
\sum\limits_{m=\hat{m}\GG+1}^\infty{\!\Prob_m\left(x\GG^2\right)}&<&\sum\limits_{m=\hat{m}\GG+1}^\infty{1-\mathrm{F}_{\chi^2\left(m\right)}\!\left(\frac{m\left(1+x\GG^2\right)}{2^{\left.2k\middle/\hat{m}\GG\right.}}\right)}\nonumber\\
&<&\sum\limits_{m=\hat{m}\GG+1}^\infty{1-\mathrm{F}_{\chi^2\left(m\right)}\!\left[m\left(1+0.5x\GG^2\right)\right]}\nonumber\\
&\leq&\sum\limits_{m=\hat{m}\GG+1}^\infty{\left[\left(1+0.5x\GG^2\right)e^{-0.5x\GG^2}\right]^{\left.m\middle/2\right.}}\label{eqn:Chernoff_Bound_Sum_Pm},
\IEEEeqnarraynumspace
\end{IEEEeqnarray}
where \eqref{eqn:Chernoff_Bound_Sum_Pm} follows from Chernoff bound for the chi-square tail probability. Since $0<\left(1+0.5x\GG^2\right)e^{-0.5x\GG^2}<1$, the sum of the geometric series in \eqref{eqn:Chernoff_Bound_Sum_Pm} is finite. The second criterion of $\hat{m}\GG$ is that \eqref{eqn:Chernoff_Bound_Sum_Pm} is upper bounded by a small $\varepsilon=10^{-2}$. Hence the error of $\tau\GG$ incurred by computing only the first $\hat{m}\GG$ terms is bounded by $\varepsilon$. The computation of $\tau\BB$ is carried out in the same manner as $\tau\GG$, but $\gamma=x\BB^2$ is used instead.

Next, we show the benefit of channel selection using pilots by comparing two schemes. The optimal selective scheme sends pilots to estimate the channel with the optimal parameter pair $\left(n^\star,T^\star\right)$ given in Section~\ref{sec:Main_Proof}. If the data transmission begins, the IR scheme mentioned earlier is used. The non-selective scheme always initiates the data transmission without estimating the channel, and the same IR scheme is used.

\begin{figure}[!t]
\centering
\includegraphics[width=0.99\columnwidth]{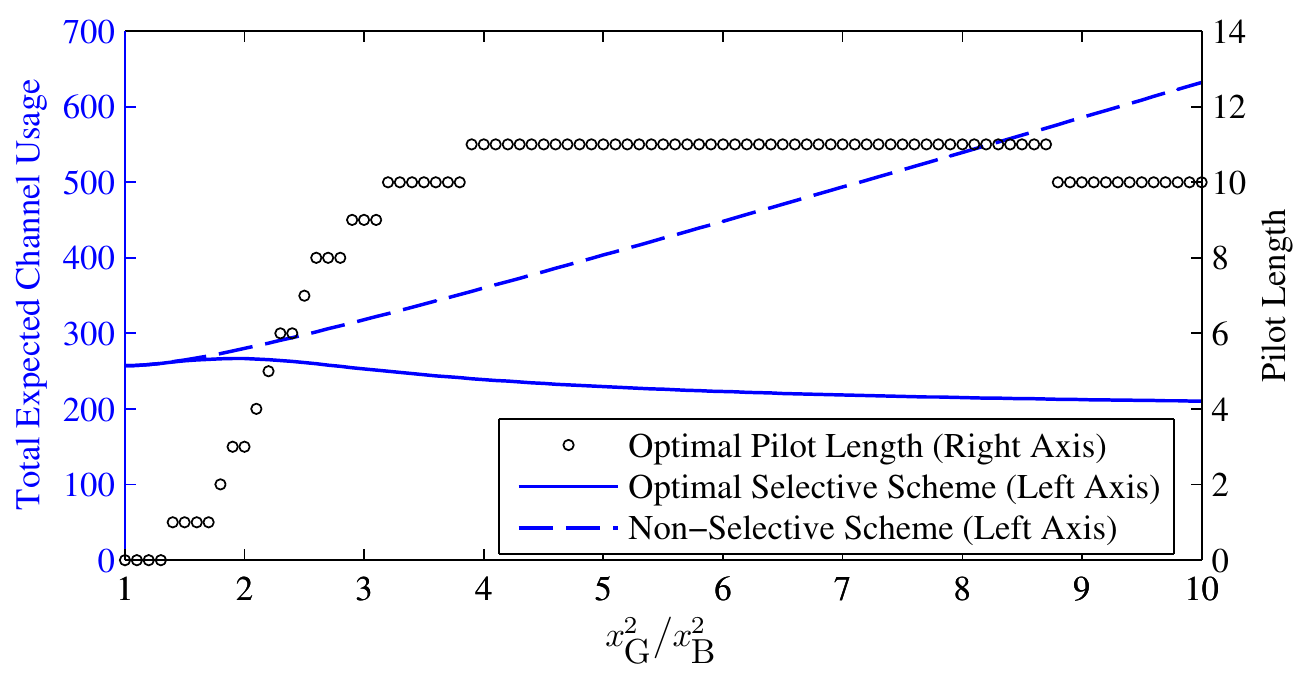}
\caption{A comparison of $\left.\bar{\tau}\right|_{n=n^\star,T=T^\star}$ and $\bar{\tau}_\textnormal{ref}$. The good and bad states are equally probable $p\GG=p\BB=0.5$, and the ratio of the associated SNRs varies while the averaged SNR is fixed to $p\GG x\GG^2+p\BB x\BB^2=1$ ($0$ dB). The packet size is $k=128$ bits.}
\label{fig:Vary_Ratio}
\end{figure}
In Fig.~\ref{fig:Vary_Ratio}, we plot the total expected channel usage (including training) of the optimal selective scheme and the non-selective scheme. The two channel states are equally probable and the averaged SNR is one. As the difference between the good and bad states increases, there is more room for the non-selective scheme to improve and smart channel selection can bring more reduction in channel usage. If the two states are similar to each other ($x\GG^2$ is close to $x\BB^2$), it requires more training symbols to distinguish the two states and the potential expected reduction $\Delta\bar{\tau}$ is small. In this case, the best choice of the selective scheme is ``no selection'' ($n^\star=0$) and the two schemes have the same performance.

\begin{figure}[!t]
\centering
\includegraphics[width=0.99\columnwidth]{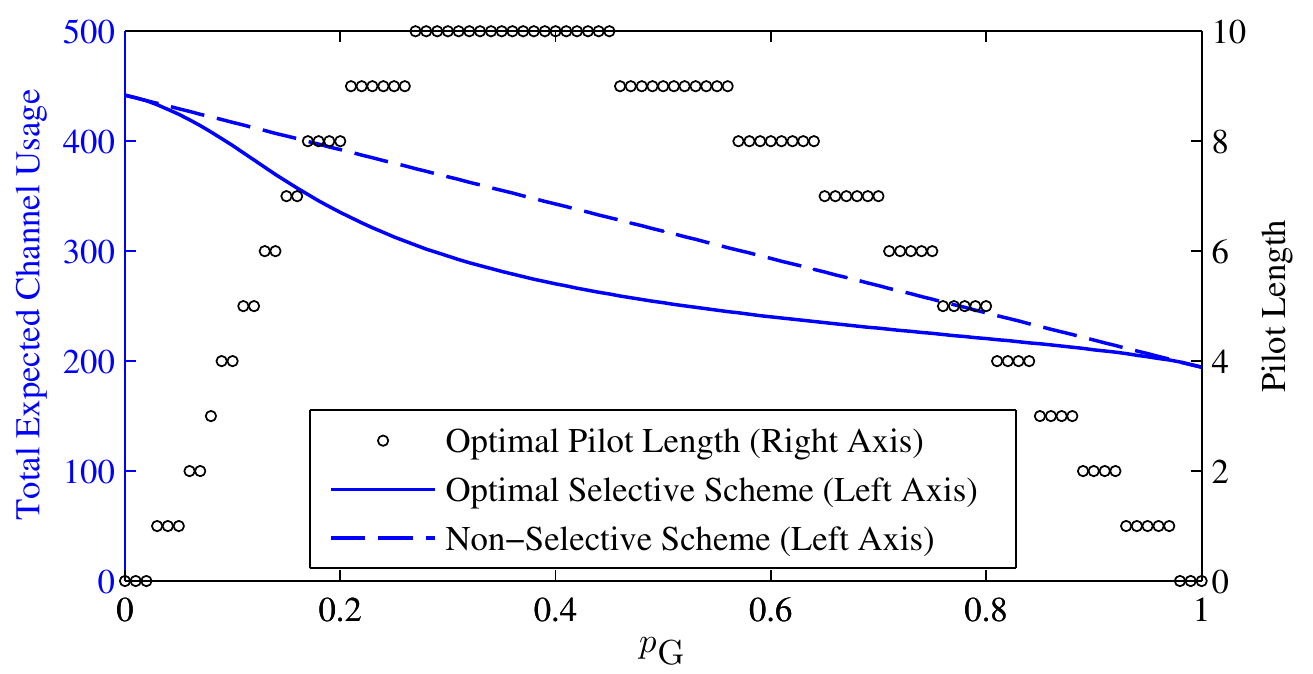}
\caption{A comparison of $\left.\bar{\tau}\right|_{n=n^\star,T=T^\star}$ and $\bar{\tau}_\textnormal{ref}$. The SNRs of the good and bad states are fixed to $x\GG^2=1.5$ ($1.76$ dB) and $x\BB^2=0.5$ ($-3.01$ dB), and the associated probabilities vary. The packet size is $k=128$ bits.}
\label{fig:Vary_Probability}
\end{figure}
A similar comparison is illustrated in Fig.~\ref{fig:Vary_Probability}. Now the probabilities of the good and bad states vary while the associated SNRs are fixed. The largest reduction appears at moderate probability. When $p\GG$ approaches zero or one, the two-state channel acts like a static channel and requires almost no training. In this case, the optimal pilot length goes to zero and the channel usage reduction go to zero.

\section{Conclusion}
\label{sec:Conclusion}
In this paper, we present a channel selection scheme based on training. After receiving several training symbols, the receiver determines whether the data transmission should be initiated or postponed. The optimal pilot length and the threshold to leave the channel balance the cost of training and the reliability of channel estimation such that the total expected channel usage is minimized. We prove that the optimal parameter pair can be achieved using Newton's method, and describe a procedure to compute it. The reduction in channel occupancy is shown for various channel settings and is evident when the two channel states are very distinct and occur with comparable probabilities.

Note that the derivation is independent of the specific transmission scheme used in our numerical results. For example, channel selection based on training could be applied to incremental redundancy with less frequent decoding attempts \cite{Adam_Limited_IR} or repeated redundancy using Chase combining \cite{Chase_Combining}. Although this paper focuses on two-state channels,  we are currently working on generalizing this concept to multiple-state channels and even continuous-coefficient block-fading channels.


\bibliographystyle{IEEEtran}
\bibliography{IEEEabrv,Pilot_Length_bib}

\end{document}